\documentclass[aps,prc,superscriptaddress,nofootinbib]{revtex4-2}

\usepackage{graphicx}
\usepackage{amsmath}
\usepackage{color}

\begin{document}


\title{Nuclear clustering process in heavy-ion collisions : experimental constraints on the low-temperature region of the QCD phase diagram}


\author{E. Bonnet}
\email{Contact author: eric.bonnet@subatech.in2p3.fr}
\affiliation{SUBATECH UMR 6457, IMT Atlantique, Université de Nantes, CNRS-IN2P3, F-44300 Nantes, France}
\author{B. Borderie}
\affiliation{Université Paris-Saclay, CNRS/IN2P3, IJCLab, 91405 Orsay, France}
\author{R. Bougault}
\affiliation{Normandie Université, ENSICAEN, UNICAEN, CNRS/IN2P3, LPC Caen, F-14000 Caen, France}
\author{A. Chbihi}
\affiliation{GANIL, CEA/DRF-CNRS/IN2P3, Bvd. Henri Becquerel, F-14076 Caen Cedex, France}
\author{Q. Fable}
\affiliation{GANIL, CEA/DRF-CNRS/IN2P3, Bvd. Henri Becquerel, F-14076 Caen Cedex, France}
\author{J.D. Frankland}
\affiliation{GANIL, CEA/DRF-CNRS/IN2P3, Bvd. Henri Becquerel, F-14076 Caen Cedex, France}
\author{D. Gruyer}
\affiliation{Normandie Université, ENSICAEN, UNICAEN, CNRS/IN2P3, LPC Caen, F-14000 Caen, France}
\author{M. La~Commara}
\affiliation{INFN, Sezione di Napoli, via Cintia, 80126 Napoli, Italy}
\affiliation{Dipartimento di Scienze Fisiche, Università di Napoli “Federico II,” via Cintia, 80126 Napoli, Italy}
\author{A.~Le~F\`{e}vre}
\affiliation{GSI Helmholtzzentrum f\"{u}r Schwerionenforschung GmbH, D-64291 Darmstadt, Germany}
\author{N. Le~Neindre}
\affiliation{Normandie Université, ENSICAEN, UNICAEN, CNRS/IN2P3, LPC Caen, F-14000 Caen, France}
\author{I. Lombardo}
\affiliation{Dipartimento di Fisica e Astronomia, University of Catania, Via Santa Sofia 64, I-95123 Catania, Italy}
\affiliation{INFN Sezione di Catania, Via Santa Sofia 64, I-95123, Catania, Italy}
\author{J.~{\L}ukasik}
\affiliation{H. Niewodnicza\'{n}ski Institute of Nuclear Physics, Pl-31342 Krak\'{o}w, Poland}
\author{M. P\^arlog}
\affiliation{Normandie Université, ENSICAEN, UNICAEN, CNRS/IN2P3, LPC Caen, F-14000 Caen, France}
\affiliation{National Institute for R\&D in Physics and Nuclear Engineering (IFIN-HH),
P. O. Box MG-6, Bucharest Magurele, Romania}
\author{G. Verde}
\affiliation{INFN Sezione di Catania, Via Santa Sofia 64, I-95123, Catania, Italy}
\affiliation{Laboratoire des 2 Infinis - Toulouse (L2IT-IN2P3), Université de Toulouse, CNRS, F-31062 Toulouse Cedex 9, France}
\author{J.P. Wieleczko}
\affiliation{GANIL, CEA/DRF-CNRS/IN2P3, Bvd. Henri Becquerel, F-14076 Caen Cedex, France}

\collaboration{INDRA Collaboration}

\date{\today}

\begin{abstract}
In this article, we study the production of hydrogen and helium isotopes in heavy-ion collisions in the incident energy range between 80 and 150 MeV/nucleon. We compare the inclusive multiplicities emitted in the transverse plane of the reaction with the predictions given by the thermal model. As a first step, we validate the choice of this approach to describe the experimental measurements. We also show that the transient states have to be explicitly taken into account for a good statistical description of the experimental multiplicities. From the thermodynamical parameter values obtained we complete the existing database built with the use of thermal-statistical models to reproduce particle production in the (ultra-)relativistic-energy measurements. We then proposed a new constraint on the so-called ``freeze-out" region in the temperature (T) \textit{versus} baryonic chemical potential ($\mu_{B}$) phase diagram of the quantum chromodynamics. These new results indicate that there is a common framework to describe the hadron production and nuclear clustering processes in heavy-ion collisions.
\end{abstract}


\maketitle

\section{Introduction}

The successful use of statistical thermal models to reproduce hadron production rates in heavy-ion collisions at (ultra-)relativistic energies has enabled us to identify the region of the quantum chromodynamics (QCD) phase diagram where hadron production rates are compatible with the chemical equilibrium hypothesis. This region can be summarized, to first order, by a so-called ``freeze-out" line in the temperature (T) versus baryonic chemical potential ($\mu_{B}$) plane and has been empirically determined from experimental data coming from several experiments carried out at different facilities~\cite{PhysRevC.73.034905}. At present, this ``freeze-out" line is only constrained in the domain $\mu_{B}\leq$800 MeV, as for values above there are no data available.
\newline
To complement the existing systematics, we propose to perform a similar analysis by comparing thermal model predictions with experimental multiplicities of hydrogen and helium isotopes collected with the INDRA detector and produced in heavy-ion collisions in the incident energy range between 80 and 150 MeV/nucleon.
In the energy domain considered, the number of particles produced in a given collision is much smaller than at relativistic energies, and does not allow a collision-by-collision statistical description. We therefore consider the set of collisions collected as a representative and unbiased statistical ensemble that can be described by global thermodynamic potentials of the thermal model. The richness and complexity of the dynamics at work in the first few moments of the collision is, of course, out of the scope using a statistical model. In stead such model allows to determine the instant for which the distribution of nucleons is supposed to not evolve anymore.
It is this instant that we are seeking to characterize, in average, in terms of thermodynamic potentials, in order to set a marker that says: whatever dynamic processes operate upstream during collisions and lead to the production of clusters, the cluster multiplicities they generate must be compatible with the prescription given by the thermal model.
\newline
In the following, the use of ``clustering process" will refer to any mechanism which ultimately produces clusters and free nucleons. No conclusions will be drawn about the nature or the dynamic of the process. Only the thermodynamic conditions with which it can be associated will be extracted.
\newline
The article is organized as follows: in the first section, we detail the data processing used to extract experimental multiplicities. In a second section, we recall the Grand-Canonical (GC) equations on which the thermal model is built, and we describe the specific implementation and corresponding ingredients. The last section presents and discusses the results obtained.

\section{Experimental details}

\begin{table}
\centering
\begin{tabular}{c|c||c|c|c|c}
\hline
Systems & $E_{beam}$ & $K^{(c.m.)}_{tot}$ & $\beta^{(lab)}_{c.m.}$ & $A_{tot}$ & $Y^{(coll)}_{p}$\\
\hline
\hline
$^{58}$Ni+$^{58}$Ni & 82 & 20.3 & 0.205 & 116 & 0.48 \\
$^{58}$Ni+$^{58}$Ni & 90 & 22.2 & 0.215 & 116 & 0.48 \\
\hline
$^{129}$Xe+$^{124}$Sn & 80 & 19.8 & 0.207 & 253 & 0.41 \\
$^{124}$Xe+$^{112}$Sn & 100 & 24.6 & 0.237 & 236 & 0.44 \\
$^{124}$Xe+$^{124}$Sn & 100 & 24.7 & 0.226 & 248 & 0.42 \\
$^{129}$Xe+$^{112}$Sn & 100 & 24.6 & 0.241 & 241 & 0.43 \\
$^{129}$Xe+$^{124}$Sn & 100 & 24.7 & 0.230 & 253 & 0.41 \\
$^{124}$Xe+$^{124}$Sn & 150 & 36.8 & 0.273 & 248 & 0.42 \\
$^{129}$Xe+$^{124}$Sn & 150 & 36.8 & 0.278 & 253 & 0.41 \\
\hline
$^{197}$Au+$^{197}$Au & 80 & 19.8 & 0.203 & 394 & 0.40 \\
$^{197}$Au+$^{197}$Au & 100 & 24.7 & 0.226 & 394 & 0.40 \\
$^{197}$Au+$^{197}$Au & 150 & 36.8 & 0.273 & 394 & 0.40 \\
\end{tabular}
\caption{Summary table of the reactions studied : incident beam energies ($E_{beam}$) and total kinetic energies in the center of mass of the reaction ($K^{(c.m.)}_{tot}$) are in MeV/nucleon; the velocity of the center of mass of the reaction ($\beta^{(lab)}_{c.m.}=p^{(lab)}_{tot}/E^{(lab)}_{tot}$) around which the transverse plane is defined; the number of nucleons ($A_{tot}$) and the proton fraction of the system ($Y^{(coll)}_{p}=Z_{tot}/A_{tot}$). Ni+Ni reactions were collected at the GANIL facility with a minimum trigger multiplicity of 4, and the others at the GSI facility with a multiplicity of 3.}
\label{tab:reactions}
\end{table}

Table~\ref{tab:reactions} lists the 12 reactions included in the analysis.
The three systems are: Ni+Ni, Xe+Sn and Au+Au for beam energies between 80 and 150 MeV/nucleon. These data have been published in a number of systematic analyses, for example, on critical phenomena~\cite{PhysRevC.71.034607} or on the stopping power of nuclear matter~\cite{PhysRevLett.104.232701}.
\newline
As the data were collected with the INDRA apparatus, we briefly recall its characteristics~\cite{POUTHAS1,POUTHAS2}. It is a 4$\pi$ multi-detector covering 90\% of solid angle around the target and is dedicated to the exclusive detection and identification of charged particles produced in heavy-ion collisions around the Fermi energy regime. It is made up of cylindrically symmetric rings surrounding the beam axis. These rings are composed of detection telescopes, the last detection stage of which (Cesium Iodide scintillator (CsI)) enables isotopes up to beryllium to be identified in atomic (Z) and mass (A) numbers. In this range, all hydrogen and helium emitted in the transverse plane reach the CsI detectors and are fully characterized. The angular range corresponding to this detection zone lies at polar angles ($\theta$) between 2 and 70 degrees with respect to the beam direction. INDRA's detection and identification performances for $^{1,2,3}$H and $^{3,4,6}$He isotopes allow us to consider the collected data as unbiased by detection.
\newline
For all the reactions, we apply the following method: we select the isotopes emitted in the transverse plane of the reaction around the velocity of the center of mass of the reaction, $\beta^{(lab)}_{c.m.}$. We define a longitudinal velocity interval $\Delta \beta_{l}=0.02$ and the yields integrated between $\beta_{l}^{-}=\beta^{(lab)}_{c.m.}-\frac{\Delta \beta_{l}}{2}$ and $\beta_{l}^{+}=\beta^{(lab)}_{c.m.}+\frac{\Delta \beta_{l}}{2}$ constitute the experimental data for the analysis.
\newline
As INDRA is a beam-on-target reaction detector, the direction of the telescopes points to the target position. The angular aperture ($\theta^{(min)}_{r}$,$\theta^{(max)}_{r}$) of the rings must be taken into account in the selection of the transverse plane: $\beta_{l}^{\pm}=\beta^{(lab)}_{c.m.}\pm\frac{\Delta \beta_{l}}{2}$. If we consider, for a given ring,  a particle with velocity ($\beta$), under the isotropic hypothesis, its real longitudinal velocity could be indiscriminately between $\beta^{(min)}_{l}=\beta cos\theta^{(max)}_{r}$ and $\beta^{(max)}_{l}=\beta cos\theta^{(min)}_{r}$.
From this uncertainty arises the probability that the particle has been emitted in the transverse plane or not. This is why we apply the following weighting to the production rates detected, $y_{r}^{(det)}(\beta)$: $y_{r}^{(exp)}(\beta)=y_{r}^{(det)}(\beta)\times\frac{\delta^{\cap}_{r}(\beta)}{\delta_{r}}$ with $\delta_{r}=\beta^{(max)}_{l}-\beta^{(min)}_{l}$ and $\delta^{\cap}_{r}(\beta)=\mathrm{min}\left(\beta^{(max)}_{l},\beta^{+}_{l}\right)-\mathrm{max}\left(\beta^{(min)}_{l},\beta^{-}_{l}\right)$. Once this weighting has been applied, the weighted contributions of the various rings are added together and integrated over the velocity range : $y^{(exp)}=\int \sum_{rings}y_{r}^{(exp)}(\beta)\mathrm{d}\beta$. The integrated yields are then normalized to the number of collisions ($N_{coll}$) collected to obtain the mean experimental multiplicities: $m^{(exp)}=\frac{1}{N_{coll}}y^{(exp)}$. These multiplicities will be compared to the prediction of the thermal model. All the multiplicity values used in the following are listed in~\cite{SuppMat}.

\section{Presentation of the thermal model}
\subsection{Grand Canonical ensemble}
The grand-canonical (GC) ensemble is used to describe a system whose number of constituents (N) can vary. This variation translates into an additional thermodynamic potential in the total energy differential: $dE=TdS+PdV+\mu dN$~\cite{LANDAU}. The chemical potential, $\mu=(dE/dN)_{S,V}$, illustrates, for the system, the energy cost associated with the variation in the number of particles. This relationship can be generalized to several types of particles with one chemical potential for each type of particle. In an ideal two-dimensional gas at temperature T, the density number n=N/V of a particle of mass M and spin s will be written according to the following equation: 
\begin{equation}
n^{(GC)}=d\int \frac{d^{2}p}{h^{2}}\left(e^{\frac{E(p)-\mu}{T}}+\kappa\right)^{-1}\label{eqn:nGC}
\end{equation}
with h is the Planck constant, $d=2s+1$ is the spin degeneracy and $\kappa$ indicates which type of statistics is followed by the particle~\footnote{If we were in the regim of Fermi-Dirac and Bose-Einstein statistics, $\kappa$=-1 or 1 will be respectively for bosons and fermions}. If the particles follow the Maxwell-Boltzmann statistics ($\kappa=0$) and the energy regime allows the single-particle energy to be approximated by $E(p)\approx p^{2}/2M+M$. Then the density number can be written as:
\begin{equation} n^{(MB)}=\frac{d}{\lambda^{2}_{T}}e^{\frac{\mu-M}{T}}\label{eqn:nMB}\end{equation}
where $\lambda_{T}=\frac{h}{\sqrt{2\pi MT}}$ is the thermal wavelength and comes from the integration of the thermal motion in momentum space : $\frac{1}{\lambda^{2}_{T}}=\int_{0}^{\infty}\frac{d^{2}p}{h^{2}}e^{-p^{2}/(2MT)}$\footnote{We perform the thermal model calculation in two dimension as we built experimental multiplicities in the transverse plane. To go from dimension 2 to 3, we simply need to increase the wavelength power from 2 to 3 as well. It is done for baryonic density and proton fraction values computed with the thermal model and used in the discussion of the last section of this article.}.

\subsection{Two-steps process calculation}
In the present analysis, we use the predictions of the thermal model that describes a two-dimensional gas of nucleons. Free nucleons (proton and neutron) and clusters, in which nucleons can be bound, are the accessible states of the system. All of these states are defined by their number of protons (Z) and neutrons (N) with A=N+Z. They are referred to in the following as ground states with $n_{gs}=n^{(MB)}$. For A$>$3 clusters, excited states are also accessible and their contribution, $\sum_{l}^{L}n_{l}$, is added to that of the ground state, with L being the number of excited states of the cluster and $n_{l}=\frac{d_{l}}{\lambda^{2}_{T}}e^{\frac{\mu-M}{T}}e^{-\frac{E_{l}}{T}}$, the contribution of a given level of energy $E_{l}$ and spin degeneracy $d_{l}$. Finally, the primary distribution of nucleons among all available states can be summarized as follows:
\begin{align}
    \left\{
      \begin{array}{lll}
        n^{(prim)}  &=n_{gs}+\sum_{l}^{L}n_{l}&\text{ for } A>3\\
                    &=n_{gs}&\text{ for } A\leq3
      \label{eqn:nprim}
      \end{array}
    \right.
\end{align}
This notion of primary distribution comes from the fact that some populated states are unstable. Nucleons populate states that will decay in very short time periods compared to detection times involved in heavy-ion collisions~\footnote{The widths of these transient states vary between eV and MeV, and are therefore neglected (set to zero) in view of the masses involved of several GeV}.
These transient states include both excited states and also the ground state of certain clusters which subsequently referred to as resonances. Their treatment require an additional step in the calculation: after the first step, where the primary distribution of nucleons among neutrons, protons and clusters is computed according to the thermal model equations~\ref{eqn:nMB}~and~\ref{eqn:nprim}, a second step is applied: we perform the redistribution of nucleons from the transient states to the stable ground states. In the end only stable states are populated~\footnote{the treatment is similar to that performed for baryonic resonances in the Resonant Hadron Gas model used at higher energies}. The associated density numbers, $n^{(sec.)}$, are written as:
\begin{equation}
n_{i}^{(sec.)} = n_{gs_{i}}+C^{(res)}_{i}+C^{(ex.st.)}_{i}\label{eqn:nsec}
\end{equation}

\begin{equation}
C^{(res)}_{i}= \sum_{j\in\mathrm{resonance}}\left(b_{gs_{j}\rightarrow gs_{i}}\times n_{gs_{j}}\right)\label{eqn:Cres} 
\end{equation}
\begin{equation}
C^{(ex.st.)}_{i}= \sum_{l_{i}=1,l_{i}\rightarrow gs_{i}}^{L_{i}}n_{i,l_{i}}+\sum_{j\neq i}\left(\sum_{l_{j}=1,l_{j}\rightarrow gs_{i}}^{L_{j}}b_{l_{j}\rightarrow gs_{i}}\times n_{j,l_{j}}\right)\label{eqn:Cexst}
\end{equation}

 In eq.~\ref{eqn:nsec}, the first term, $n_{gs}$, is the ground state density number of the cluster, the second term ($C^{(res)}$, eq.~\ref{eqn:Cres}) is the contribution of the resonances that populate this cluster by their decay and the last term ($C^{(ex.st.)}$, eq.~\ref{eqn:Cexst}) is of the excited states that also populate this cluster. $C^{(ex.st.)}$ includes two contributions: the first one is from excited states of this cluster whose energies are below the particle emission threshold and which decay by isomeric transition, and the second one is from excited states of other clusters. In $C^{(res)}$ and $C^{(ex.st.)}$ terms, the branching ratios ($b_{j\rightarrow i}$) illustrate the contribution of decays towards stable states~\footnote{For explanation purpose, if we take the example of the decay of the $^{5}$He resonance, we get, for its ground state $b_{^{5}He\rightarrow^{4}He}=1$ and $b_{^{5}He\rightarrow^{1}n}=1$. For $^{8}$Be ground state, we get : $b_{^{8}Be\rightarrow^{4}He}=2$}. Note that the transition between both steps, described above, preserves the total baryonic density ($n_{B}=\sum_{i} A_{i}n_{i}$) and the proton fraction of the system ($Y_{p}=(\sum_{i} Z_{i}n_{i}) /n_{B}$).


\subsection{Ingredients, hypothesis and fit procedure\label{subsection}}
We have included in the calculation neutron and all isotopes from hydrogen to beryllium (34 ground states) and 190 excited states. Only states with known spin have been taken into account which correspond to around 70\% of the measured levels listed. It should be noted that our knowledge of excited states is incomplete and is likely to be supplemented by future measurements~\footnote{
All data comes from the IAEA database accessible on the website https://www-nds.iaea.org/relnsd/vcharthtml/VChartHTML.html}.
To summarize, we will distinguish between 19 stable states ($^{1}$n, $^{1,2,3}$H, $^{3,4,6,8}$He, $^{6,7,8,9,11}$Li, $^{7,9,10,11,12,14}$Be) among which $^{1,2,3}$H and $^{3,4,6}$He are the ones collected experimentally and used for comparison and 15 resonances ($^{4,5}$H, $^{5,7,9,10}$He, $^{4,5,10,12}$Li, $^{6,8,13,15,16}$Be).
\newline
In order to compare the thermal model with the experimental data, it is essential to propose a relationship to reduce the number of unknowns, namely, the 34 chemical potentials associated with free nucleons and clusters. The simplest and most common way to do this is to consider the chemical equilibrium of the system. This equilibrium implies that, for all particles, the decomposition of the chemical potential is related to the conserved quantities of the system, in this case, the baryonic number (B=A=N+Z) and the electric charge (Q=Z). This results in the following decomposition:
\begin{equation}\mu_{i}=\mu(Z_{i},A_{i})=A_{i}\mu_{B}+Z_{i}\mu_{Q}\label{eqn:mu}\end{equation}
This allows us to calculate density numbers for all accessible states with just the 3 following parameters: temperature (T), baryonic chemical potential ($\mu_{B}$) and electrical potential ($\mu_{Q}$).
\newline
The last step to compare with the experimental multiplicities is to impose the conservation of the number of nucleons with the following relation involving a normalization volume (V):
\begin{equation}
m_{i}^{(fit)}=n^{(sec.)}_{i}\times V\label{eqn:mfit}
\end{equation}
with
\begin{equation}
V=(\sum_{i}A_{i}\times m_{i}^{(exp)})/(\sum_{i}A_{i}\times n_{i}^{(sec.)})\label{eqn:V}
\end{equation}

Concerning the $\mu_{Q}$ potential, in high-energy analysis, its value is most often deduced by making the hypothesis of conservation of the proton fraction of the colliding system ($Y_{p}^{(coll)}$). We have chosen to leave it free, since there is nothing \textit{a priori} to assure us that the repartition between neutrons and protons remains fixed and equal to $Y_{p}^{(coll)}$ during the collision in a thin slice of longitudinal velocity that is the transverse plane.
\newline
Finally, we perform a three-parameter fit (T, $\mu_{B}$ and $\mu_{Q}$) to make the multiplicities $m_{i}^{(exp)}$ and $m_{i}^{(fit)}$ of $^{1,2,3}$H and $^{3,4,6}$He coincide as much as possible. For minimization, we use the MINUIT~\cite{Minuit} package implemented in the ROOT software using nucleon-sharing reproduction (NSR) as the estimator defined in eq.~\ref{eqn:NSR}.
\begin{equation}
\mathrm{NSR}=1-\frac{\sum_{i}A_{i}|m^{(fit)}_{i}-m^{(exp)}_{i}|}{\sum_{i}A_{i}m^{(exp)}_{i}}
\label{eqn:NSR}
\end{equation}
To better understand the different effects of the transient states in the final stable state density number (eq.~\ref{eqn:nsec}), we performed the fitting procedure by adding them successively to see their effects, firstly, on the ability to reproduce the data and secondly on the parameter values obtained.
We have performed the fit procedure for these three cases:
\begin{itemize}
    \item case 1, no transient states are populated and therefore do not contribute to final stable states : $C^{(res)}=0$ and $C^{(ex.st.)}=0$ in eq.~\ref{eqn:nsec}.
    \item case 2, only resonances are populated and contribute to final stable states : $C^{(res)}\neq0$ and $C^{(ex.st.)}=0$ in eq.~\ref{eqn:nsec}.
    \item case 3, resonances and excited states are populated and contribute to final stable states : $C^{(res)}\neq0$ and $C^{(ex.st.)}\neq0$ in eq.~\ref{eqn:nsec}.
\end{itemize}

\section{Results and discussion}

\subsection{Data reproduction}

\begin{figure}
\includegraphics[width=1.0\textwidth]{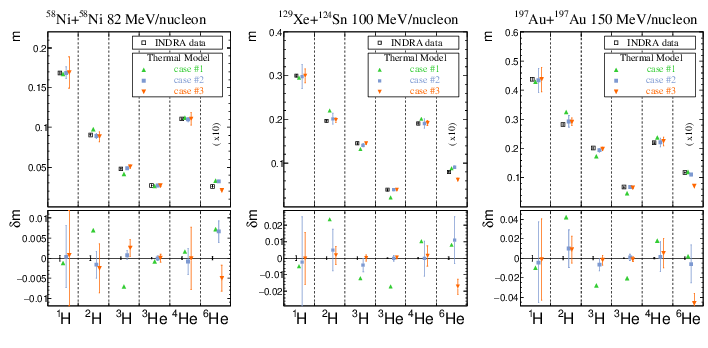}
\caption{\label{fig:fig01} Comparison of hydrogen and helium isotope multiplicities; the three panels correspond to the three reactions studied in this work. The upper part of the panels shows a comparison between the experimental multiplicities (black open squares, $m^{(exp)}$) and those obtained by the thermal model (solid colored symbols, $m^{(fit)}$). The lower part of each panel shows the differences between thermal model results and experimental data ($\delta m$). For the thermal model, the three colors correspond to the three cases described in the text and show the evolution in data reproduction of the inclusion of resonances (case 2, blue solid squares), the inclusion of resonances and excited states (case 3, orange inverted solid triangles) compared with case 1, where only the ground states of stable isotopes are accessible (green solid triangles). For better visibility, $^{6}$He multiplicities are multiplied by 10. For cases 2 and 3, the error bars for the multiplicities obtained with the model are deduced from the uncertainties obtained for the parameters (T, $\mu_{B}$ and $\mu_{Q}$) at the end of the fit procedure. For case 1, where the fit does not converge, we have not reported these error bars. On the lower panels, experimental errors are also indicated by black lines.
}
\end{figure}

The experimental multiplicities (open black squares) for three of the reactions studied are shown on top panels of figure~\ref{fig:fig01}. To these are added the different predictions of the thermal model according to the different cases listed before (full colored symbols). For easier comparison, on bottom panels, we display the difference between prediction and experimental points, $\delta m = m^{(fit)}-m^{(exp)}$. To correctly reproduce the data, the introduction of transient states is necessary. For case~1, where they are not included, we observe discrepancies in the prediction of multiplicities and, specially, the right balance between the $^{2,3}$ H and $^{3}$ He isotopes is not achieved and this is more pronounced for heavier systems and higher bombarding energies. For the predictions of cases 2 and 3, there are no noticable differences, and we observe a very good agreement with the data. We thus confirm the need to take transient states into account in this type of comparison, as already mentioned in the conclusion of~\cite{VOVCHENKO2020135746}. To be quantitative, the NSR estimator (eq.~\ref{eqn:NSR}) is above 97\% for these two latter cases for all reactions included in this comparison while, for case~1, the NSR is around 95\% for Ni+Ni reactions and drops to only 90\% for Xe+Sn et Au+Au reactions. It should be noted that the $^{6}$He multiplicities are slightly less well reproduced.
The contribution of transient states, $P_{tr.st.}$, populated during the first step is computed with eq.~\ref{eqn:Ares}. For case~2, the percentage of nucleons that are redistributed is between 21 and 28\% while for case~3, values are between 35 and 45\%.

\begin{equation}
P_{tr.st.}=\sum_{i\in stable} A_{i}\left(n_{i}^{(sec.)}-n_{gs_{i}}\right) / \sum_{i\in stable} A_{i}n_{i}^{(sec)}
\label{eqn:Ares}
\end{equation}

In the following discussion of the values obtained for the fit parameters, we will focus only on cases 2 and 3 for which the thermal model reproduce experimental data.

\subsection{Parameter trends}

\begin{table}
    \centering
\begin{tabular}{c|c||c|c|c|c||c|c|c|c}
\hline
\multicolumn{2}{c||}{Reactions} & \multicolumn{4}{c||}{case 2}& \multicolumn{4}{c}{case 3}\\
\hline
Systems & $E_{beam}$  & T [MeV]& $\mu_{B}$ [MeV]& $\mu_{Q}$ [MeV]& $n_{B}$[nuc.fm$^{-3}$]  & T [MeV]& $\mu_{B}$ [MeV]& $\mu_{Q}$ [MeV]& $n_{B}$[nuc.fm$^{-3}$] \\
\hline
$^{58}$Ni+$^{58}$Ni & 82 &6.5$\pm$0.1 & 927.1$\pm$0.1 & -3.5$\pm$0.2 & 0.015$\pm$0.002& 	5.7$\pm$0.1 & 926.9$\pm$0.2 & -3.1$\pm$0.4 & 0.010$\pm$0.003\\
$^{58}$Ni+$^{58}$Ni & 90 &6.7$\pm$0.1 & 926.8$\pm$0.1 & -3.7$\pm$0.2 & 0.014$\pm$0.002& 	5.9$\pm$0.1 & 926.6$\pm$0.2 & -3.2$\pm$0.4 & 0.010$\pm$0.003\\
\hline
$^{129}$Xe+$^{124}$Sn & 80 &6.0$\pm$0.3 & 929.2$\pm$0.3 & -7.3$\pm$0.4 & 0.010$\pm$0.004&	5.3$\pm$0.2 & 928.9$\pm$0.3 & -7.2$\pm$0.4 & 0.006$\pm$0.002\\
$^{124}$Xe+$^{112}$Sn & 100 &6.6$\pm$0.2 & 927.6$\pm$0.2 & -5.8$\pm$0.2 & 0.012$\pm$0.002& 5.9$\pm$0.0 & 927.4$\pm$0.1 & -5.5$\pm$0.1 & 0.008$\pm$0.001\\
$^{124}$Xe+$^{124}$Sn & 100 &6.7$\pm$0.3 & 928.0$\pm$0.3 & -6.9$\pm$0.3 & 0.012$\pm$0.004& 5.9$\pm$0.1 & 927.9$\pm$0.1 & -6.8$\pm$0.2 & 0.007$\pm$0.001\\
$^{129}$Xe+$^{112}$Sn & 100 &6.6$\pm$0.2 & 927.8$\pm$0.2 & -6.2$\pm$0.3 & 0.012$\pm$0.003& 5.9$\pm$0.1 & 927.6$\pm$0.1 & -6.0$\pm$0.1 & 0.008$\pm$0.001\\
$^{129}$Xe+$^{124}$Sn & 100 &6.6$\pm$0.2 & 928.2$\pm$0.2 & -7.3$\pm$0.3 & 0.011$\pm$0.003& 5.9$\pm$0.1 & 928.0$\pm$0.1 & -7.1$\pm$0.1 & 0.007$\pm$0.001\\
$^{124}$Xe+$^{124}$Sn & 150 &7.9$\pm$0.5 & 925.7$\pm$0.6 & -6.5$\pm$0.6 & 0.012$\pm$0.006& 6.7$\pm$0.1 & 925.9$\pm$0.2 & -6.3$\pm$0.3 & 0.008$\pm$0.002\\
$^{129}$Xe+$^{124}$Sn & 150 &7.8$\pm$0.3 & 925.9$\pm$0.4 & -6.8$\pm$0.5 & 0.012$\pm$0.004& 6.7$\pm$0.1 & 926.2$\pm$0.2 & -6.7$\pm$0.3 & 0.007$\pm$0.002\\
\hline
$^{197}$Au+$^{197}$Au & 80 &6.1$\pm$0.3 & 929.4$\pm$0.3 & -7.5$\pm$0.4 & 0.013$\pm$0.005&	5.4$\pm$0.1 & 929.1$\pm$0.2 & -7.5$\pm$0.3 & 0.008$\pm$0.002\\
$^{197}$Au+$^{197}$Au & 100 &6.9$\pm$0.2 & 928.4$\pm$0.2 & -7.4$\pm$0.3 & 0.016$\pm$0.004& 5.9$\pm$0.3 & 928.4$\pm$0.4 & -7.4$\pm$0.5 & 0.009$\pm$0.005\\
$^{197}$Au+$^{197}$Au & 150 &7.9$\pm$0.3 & 926.5$\pm$0.3 & -6.8$\pm$0.4 & 0.018$\pm$0.005& 6.7$\pm$0.1 & 926.6$\pm$0.2 & -6.6$\pm$0.3 & 0.010$\pm$0.002\\
\end{tabular}
\caption{ \label{tab:parameters}For each reaction included in the analysis, values and uncertainties of the three thermal model parameters (T, $\mu_{B}$ and $\mu_{Q}$ in MeV) obtain from the fit procedure for the case 2 and 3 (see text for details). Values of deduced baryonic densities ($n_{B}=\sum_{i} A_{i}n_{i}$) are also listed. The uncertainties are obtained by independently exploring the uncertainties of the T, $\mu_{B}$ and $\mu_{Q}$ parameters. The values displayed in figures~\ref{fig:fig02} and~\ref{fig:new_potentials} are the average of the values for the two cases.}
\end{table}

In table~\ref{tab:parameters}, the values of the thermodynamic potentials obtained with the fit procedure are given with their uncertainties. The trends of the parameters are consistent with those expected as a function of incident energy ($E_{inc}$) and the proton fraction of the colliding system ($Y^{(coll)}_{p}$): temperature (T) increases and the baryonic potential ($\mu_{B}$) decreases with increasing incident energy, and the negative correction provided by the electric potential ($\mu_{Q}$) increases when the proton fraction of the system decreases. The values obtained in the model for the proton fractions vary by around plus or minus 5\% compared to the values of the colliding system listed in tab.~\ref{tab:reactions}. All of these statements are valid for both case~2 and 3.
\newline
Now, if we look in average at the differences in parameter values between case~2 and 3, we mainly observe a negative temperature shift of around $\Delta^{(3-2)} T\approx-0.87$ MeV when excited states are included. For chemical potentials, the values are compatible if we take into account the error bars of the fit ($\Delta^{(3-2)} \mu_{B}\approx-0.08$ and $\Delta^{(3-2)} \mu_{Q}\approx0.17$ MeV). 
This systematic difference on temperatures is explained by the opening of accessible states. Indeed, for the same set of parameters, the total density will be greater in case 3 (by around 30\% on average). But as the number of nucleons is constrained in the fit (eq.~\ref{eqn:V}), it is the temperature that decreases to counterbalance this opening to excited states without affecting chemical potentials. A similar shift to lower densities are consequently observed for the case~3.
\newline
In the following, the experimental data used and displayed will be the averages (black symbols) and the dispersions (error bars) of the values obtained from the two cases and listed in table~\ref{tab:parameters}.

\begin{figure}
\includegraphics[width=1.0\textwidth]{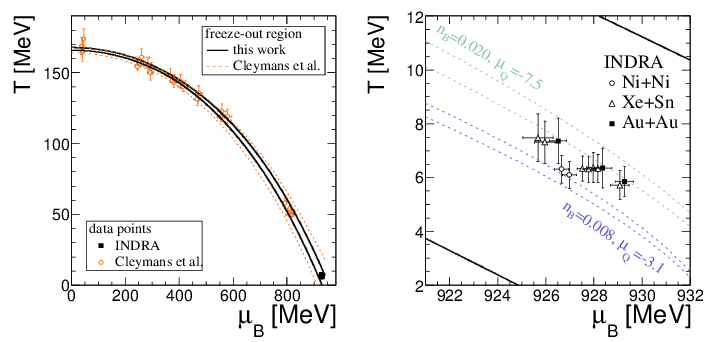}
\caption{\label{fig:fig02}Graphical representation of the parameter values adapted from those listed in Table~\ref{tab:parameters} (see text for details). Left panel, temperature (T) - baryonic chemical potential ($\mu_{B}$) phase diagram of the QCD; effects of taking INDRA data into account in the ``freeze-out" line constraint. The orange open circles are the points used in~\cite{PhysRevC.71.034607} to obtain the parameterization ($T(\mu_{B})$, eq.~\ref{eqn:TMuB}). The solid black squares correspond to the points obtained in this work. Orange dotted lines and black solid lines delimit the ``freeze-out" region before and after INDRA points are taken into account in the systematics. These regions are obtained from the parameter uncertainties resulting from the fit procedure. Right panel, zoomed in on the region populated by the present analysis: the black lines are the same as for the left panel, for easier reference. The various markers represent the three systems studied (Ni+Ni, Xe+Sn and Au+Au). The dotted lines are examples of evolution in the diagram for given values of baryonic density ($n_{B}$ in $fm^{-3}$) and chemical electric potential ($\mu_{Q}$ in MeV). The chosen values correspond to the extrema obtained with the fit procedure. As discussed in the text, the two curves of the same color and associated with the same pair of values ($n_{B}$,$\mu_{Q}$) represent the dispersion resulting from the inclusion (case 3, lower line) or exclusion (case 2, upper line) of excited states in the primary nucleon distribution among free nucleons and clusters.}
\end{figure}

\subsection{New constraints on the ``freeze-out" line}

As mentioned in the introduction, one of the motivations of this work is to provide experimental information on the localization of the nuclear clustering process in the QCD phase diagram, and to put it into perspective with existing higher-energy systematics. 
The two panels in Figure 2 are the usual representation of the QCD phase diagram in the $T-\mu_{B}$ plane. The panel on the right zooms in on the region populated by the present study. The panel on the left brings together the data used in~\cite{PhysRevC.73.034905} (open orange symbols) and data from this work (solid black symbols).  In~\cite{PhysRevC.73.034905}, the authors proposed the following parameterization to describe the observed trend:
\begin{equation}T(\mu_{B})=a-b\mu^{2}_{B}-c\mu^{4}_{B}\label{eqn:TMuB}\end{equation}
with associated parameter values obtained from the reproduction of orange symbols.
\begin{equation*}
a=0.166^{\pm0.002} \text{GeV},\;b=0.139^{\pm0.016} \text{GeV}^{-1}\text{ and }c=0.053^{\pm0.021} \text{GeV}^{-3}.
\end{equation*}
Adding our 12 new points to the systematic, we proceed in the same way performing the fit procedure using eq.~\ref{eqn:TMuB} to obtain the new values of the parameters:
\begin{equation*}
a=0.167^{\pm0.001} \text{GeV},\;b=0.135^{\pm0.007} \text{GeV}^{-1}\text{ and }c=0.060^{\pm0.007} \text{GeV}^{-3}.
\end{equation*}
For better visibility, we have chosen to draw the two edges of the ``freeze-out" region (dotted orange and full black lines) rather than just the line given by the parameter values, as it is the extent of this region that is most constrained by the new fit. This region is obtained by independently exploring the uncertainties of the a, b and c parameters. The new set of parameters obtained is compatible with the previous one with uncertainties reduced by a factor of 2 to 3. As expected, the most affected parameter is c with a modification of around 12\% of its original values. The reduction of uncertainties leading to the shrinkage of the ``freeze-out" region is most marked in the $\mu_{B}>800$ range. In addition to the thermal model's success in reproducing all the data from particle production in heavy-ion collisions, whatever the mechanism at work (from hadronization to clustering processes), this new delimitation of the ``freeze-out" region of the QCD phase diagram could be valuable for current attempts, involving microscopic approaches, aiming to follow the $T(\mu_{B})$ parameterization to extract more physical outputs (see, for example,~\cite{BLASCHKE2025139206} for comparisons and discussion).
\newline
The right panel of fig.~\ref{fig:fig02} is more focused and information related to the nuclear clustering we can extract from the use  of the thermal model. The black symbols and lines are the same as in the left panel. We have also drawn lines framing the points and symbolize the trajectories for two sets of constant values of baryonic density, $n_{B}$, and chemical electric potential, $\mu_{Q}$. These values are the extreme ones obtained from the fit procedure (table~\ref{tab:parameters}). For each set of parameters, two lines are drawn illustrating the discussion regarding the inclusion (lower line) or exclusion (upper one) of the excited states in the thermal model\footnote{For comparison, the freeze-out line is well reproduced by a constant density line ($n_{B}+n_{\bar{B}}\approx$ 0.12 fm$^{-3}$) for $\mu_{B}$ values lower than 600 MeV~\cite{PhysRevC.73.034905}.}.
\newline
Now, if we look at the density values deduced from solutions obtained with the thermal model, they indicate that the reproduction of experimental clusters multiplicities with an ideal gas, made of free nucleons and clusters at chemical equilibrium, is associated with low density values of one order of magnitude below the normal density. They are also qualitatively compatible with those obtained in cluster analyses performed on data collected at Fermi energies~\cite{Kowalski2007,Qin2012,Mabiala2014,Bougault2020}.
There are, however, differences in the computation of density between the two approaches. In our case, this density is obtained by taking into account all the states accessible to nucleons (list given in the subsection~\ref{subsection}) and deduced from the thermal model after ensuring that the data are reproduced and compatible with equilibrium. For the analyses mentioned, densities are obtained using Albergo's formulation~\cite{Albergo} or dedicated derivations. This formulation postulates an \textit{a-priori} chemical equilibrium to simplify equation~\ref{eqn:nMB}) and derive density and temperature expressions in terms of isotopic ratios\footnote{Albergo's formulation also makes the following approximation for cluster masses: $M_{i}=A_{i}\times931.49$, so that only binding energies come into play. However, if we wish to extract chemical potential values, this approximation is not suited.}. It also makes the strong assumption of a truncation of the set of accessible states: only free nucleons and stable clusters (generally $^{2,3}$H and $^{3,4}$He) come into play during the clusterization process, and no de-excitation process of transient states exists or participates in the population of these clusters.
In~\cite{Ropke2013}, a variable is proposed ($R_{test}$) to quantify the deviation of the data from the case of chemical equilibrium corresponding to $R_{test}$=1. However, this variable is also based on this truncation of accessible states. Indeed, if we calculate $R_{test}$ from equation~\ref{eqn:mfit}, with the parameter values obtained in this work (table~\ref{tab:parameters}), we obtain $R_{test}$ values ranging from 1.29 to 1.38. This confirms that, depending on the accessible states included in the calculation, conclusions about the interpretation of experimental data can change. To sum up, it would be interesting to apply our method to the experimental multiplicities used for these published low-energy analyses, to see whether taking transient states into account can reproduce the data. For the moment, these data are reproduced with assumptions of in-medium effects implemented in various ways in different models~\cite{Ropke2013,Pais2020}. It could be instructive, for example, to relate our $\mu_{Q}$ values to the Coulomb effects mentioned in~\cite{Mabiala2014}. Concerning the possible comparison of our results with predictions coming from models using baryonic density as main control parameter, the list of particles and available states used to build this density should be known and similar to allow a pertinent comparison.

\subsection{Rewriting chemical potentials}

\begin{figure}
\includegraphics[width=1.0\textwidth]{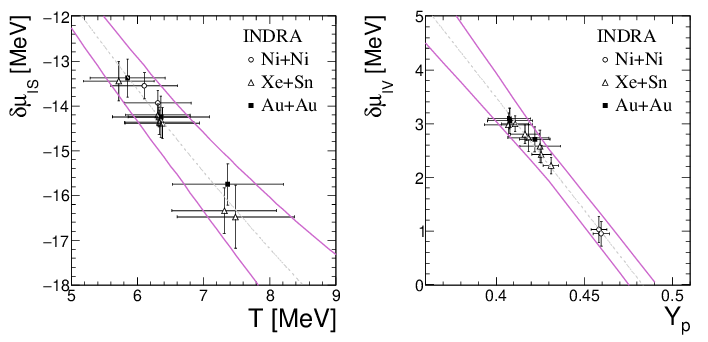}
\caption{\label{fig:new_potentials}Left panel: evolution of isoscalar chemical potential ($\delta\mu_{IS}$) as a function of temperature (T); right panel: evolution of isovector chemical potential ($\delta\mu_{IV}$) as a function of proton fraction ($Y_{p}$). The different markers represent the three systems studied (Ni+Ni, Xe+Sn and Au+Au). For both panels, the dotted line represents the interpolation proposed and the solid magenta lines show the region compatible with the errors on the parameters used in these interpolations.
}
\end{figure}

We see also in the right panel of fig.~\ref{fig:fig02} that the knowledge of only the temperature and the baryonic chemical potential is not enough to ensure a one to one relation with a given set of multiplicities. The electric charge potential is needed to remove the degeneracy. In other words, at least in this region, the isospin degree of freedom (relative proportion of neutrons and protons) plays an important role. As the definition of chemical potential written in eq.~\ref{eqn:mu} somehow hides the specificity of this neutron/proton mixture of nuclear matter, we can rewrite the chemical potential as follows: $\mu-M=A\mu_{IS}+(N-Z)\mu_{IV}-M$ with $\mu_{IS}=\mu_{B}+\frac{1}{2}\mu_{Q}$, $\mu_{IV}=-\frac{1}{2}\mu_{Q}$. This decomposition into isoscalar and isovector terms is inspired by the usual decomposition of the nuclear equation of state according to the isospin degree of freedom. To examine the behavior of these two components, we also avoid the trivial effect of the masses of the proton and neutron by looking at the evolution of $\delta\mu_{IS}=\mu_{IS}-\frac{1}{2}(M_{n}+M_{p})$ as a function of temperature and $\delta\mu_{IV}=\mu_{IV}-\frac{1}{2}(M_{n}-M_{p})$ as a function of the proton fraction. These two trends are plotted on the two panels of fig.~\ref{fig:new_potentials}. Generally speaking, the different points show a common trend and this makes it possible to propose two simple parameterizations of the chemical potentials more specific to this energy regime: $\delta\mu_{IS}(T)=-2.63^{\pm0.13}T+0.06^{\pm0.02}T^{2}$ and $\delta\mu_{IV}(Y_{p})=20.28^{\pm2.62}(1-2.07^{\pm0.03}Y_{p})$.
\newline
Since, to our knowledge, there are no published data on the evolution of chemical potentials in this temperature range, these parameterizations are a first attempt and should be completed with already published data with the knowledge of the $\mu_{Q}$ values.  In particular, it will be very interesting to get values of $\mu_{Q}$ associated to existing points around $\mu_{B}$=800 MeV (labeled SIS in~\cite{PhysRevC.73.034905}) which are the closest points to our data. From a general point of view, if experimental particle multiplicity data are available, extraction of thermodynamical potentials by means of statistical model comparison in the range $\mu_{B}$=800-900 MeV will be very stimulating.

\section{Conclusion}

In this paper, we have studied the production of hydrogen and helium isotopes in heavy-ion collisions of Ni+Ni, Xe+Sn and Au+Au for incident energies between 80 and 150 MeV/nucleon. The average multiplicities emitted in the transverse plane of the collision are reproduced by the thermal model describing an ideal gas of free neutrons and protons in chemical equilibrium with clusters. The thermodynamic potentials obtained are consistent with existing results at higher energies and validate the continuity of a universal curve in the QCD phase diagram called the ``freeze-out" line linking hadron production and nuclear clustering processes. 
With regard to the information obtained, it would be interesting to be able to compare the evolutions obtained for the isoscalar and isovector chemical potentials with those obtained in the modeling of the generalized equations of state that incorporate clusters~\cite{compOSE} to possibly establish bridges between the experimental data and the microscopic ingredients of these theoretical descriptions. The publication of $\mu_{Q}$ values corresponding to existing data could be useful and should be generalized.

\begin{acknowledgments}
This work was financed by the Centre National de la Recherche Scientifique (CNRS). The original version of this article was written in French and translated into English with the help of DeepL. 
\end{acknowledgments}

\bibliography{biblio}

\end{document}